# Unsupervised Full-color Cellular Image Reconstruction through Disordered Optical Fiber

Xiaowen Hu[1], Jian Zhao[2,*], Jose Enrique Antonio-Lopez[1], Rodrigo Amezcua Correa[1], and Axel Schülzgen[1]

[1]CREOL, The College of Optics and Photonics, University of Central Florida, Orlando, FL 32816, USA
[2]The Picower Institute for Learning and Memory, Massachusetts Institute of Technology, Cambridge, Massachusetts 02139, USA
*Corresponding authors: jianzhao@knights.ucf.edu

**Abstract**

Recent years have witnessed the tremendous development of fusing fiber-optic imaging with supervised deep learning to enable high-quality imaging of hard-to-reach areas. Nevertheless, the supervised deep learning method imposes strict constraints on fiber-optic imaging systems, where the input objects and the fiber outputs have to be collected in pairs. To unleash the full potential of fiber-optic imaging, unsupervised image reconstruction is in demand. Unfortunately, neither optical fiber bundles nor multimode fibers can achieve a point-to-point transmission of the object with a high sampling density, as is a prerequisite for unsupervised image reconstruction. The recently proposed disordered fibers offer a new solution based on the transverse Anderson localization. Here, we demonstrate unsupervised full-color imaging with a cellular resolution through a meter-long disordered fiber in both transmission and reflection modes. The unsupervised image reconstruction consists of two stages. In the first stage, we perform a pixel-wise standardization on the fiber outputs using the statistics of the objects. In the second stage, we recover the fine details of the reconstructions through a generative adversarial network. Unsupervised image reconstruction does not need paired images, enabling a much more flexible calibration under various conditions. Our new solution achieves full-color high-fidelity cell imaging within a working distance of at least 4 mm by only collecting the fiber outputs after an initial calibration. High imaging robustness is also demonstrated when the disordered fiber is bent with a central angle of 60°. Moreover, the cross-domain generality on unseen objects is shown to be enhanced with a diversified object set.

**Introduction**

Optical fibers are well-known for transmitting remote information out of otherwise inaccessible areas. Because of their miniature sizes and flexibility, fiber-optic imaging systems (FOISs) [1] have become an indispensable tool in clinical practice and biological research, such as early detection of gastrointestinal cancers [2, 3, 4, 5] and visualization of neuronal activities in freely moving animals [6, 7, 8, 9, 10]. Most common FOISs are based on optical fiber bundles or multimode fiber (MMF). An optical fiber bundle consists of thousands of closely spaced cores in a shared cladding. Each core acts as a single-pixel detector to sample the object [6, 11, 12, 13, 14]. Due to the loss of information in the cladding, the output images from an optical fiber bundle suffer from the honeycomb effect [15]. On the other hand, an MMF supports thousands of optical modes in a single core. Because of the mode coupling in MMF, object images are scrambled into speckle patterns. Supervised deep learning [16, 17] has been successfully implemented in both cases to reconstruct high-quality images [18, 19, 20, 21]. A Convolutional Neural Network (CNN) can "learn" an image reconstruction mapping from numerous pairs of fiber output images and the input object images. Despite its success, supervised deep learning imposes a heavy burden on FOISs. The collection of paired fiber outputs and input objects in the calibration step involves time-consuming and demanding alignments of the FOIS. Especially, a re-calibration is required for any system variations, which is infeasible for practical applications.

Unsupervised deep learning circumvents these hurdles by using unpaired training image data. Since the deep learning model has to uncover the hidden mapping between two image domains without paired images, image reconstruction using unsupervised deep learning is considered to be a challenging task. Recently, it has been demonstrated that if the two image domains are similar in the high-dimensional space, "generator" CNNs and "discriminator" CNNs can compete in adversarial games to find a "natural" translation between the two image domains [22, 23]. To achieve this similarity in the high-dimension domain, the FOISs should have a point-to-point transmission between the input object and the fiber output with high sampling densities. Unfortunately, neither optical fiber bundles nor MMFs meet these requirements. Although optical fiber bundles can directly convey the images of the objects, they have limited sampling densities (~0.1 mode/$\mu m^2$). As more sampling points, i.e., more cores, are added, the core-to-core crosstalk becomes stronger and degrades the point-to-point transmission fidelity [24]. On the other hand, the input-output relationship in MMFs is far deviated from a point-to-point transmission due to the multimode interference. The recently proposed glass-air Anderson localizing optical fibers (GALOFs) [25, 26, 27, 28, 29, 30, 31, 32, 33] provide a promising alternative. With a disordered arrangement of air holes embedded in a silica matrix, GALOFs achieve local confinement of light and high sampling densities (~10 mode/$\mu m^2$) simultaneously [34] due to the transverse Anderson localization (TAL) [35, 36]. Moreover, the TAL-supported modes are insensitive to external perturbations [37] or wavelength shifts [38], as opposed to the modes in optical fiber bundles[24, 39, 40] or MMFs[41, 42, 43, 44, 45]. Therefore, robust full-color image transport can be achieved.

Here, we demonstrate unsupervised full-color high-fidelity image reconstruction through a meter-long GALOF in both transmission and reflection modes. We show that a simple histogram equalization step is adequate to reveal the hidden objects in the GALOF outputs preliminarily due to the densely-distributed TAL modes. The objects' fine details can be further recovered by utilizing the unpaired image-to-image translation [22, 23]. Unsupervised image reconstruction significantly simplifies the calibration step, where the object images only need to be collected once. Therefore, the GALOF-based FOIS is flexible towards different conditions. As a remarkable example, we show the system's consistent imaging performance within a working distance of at least 4 mm, with a simple one-step re-calibration that only requires GALOF outputs. Moreover, due to the robustness of the TAL-supported modes, high image quality is preserved even under substantial mechanical bending (~60° bent angle). Finally, we show that the cross-domain generalizability of unseen objects can be enhanced by increasing the objects' diversity.

**Principles**

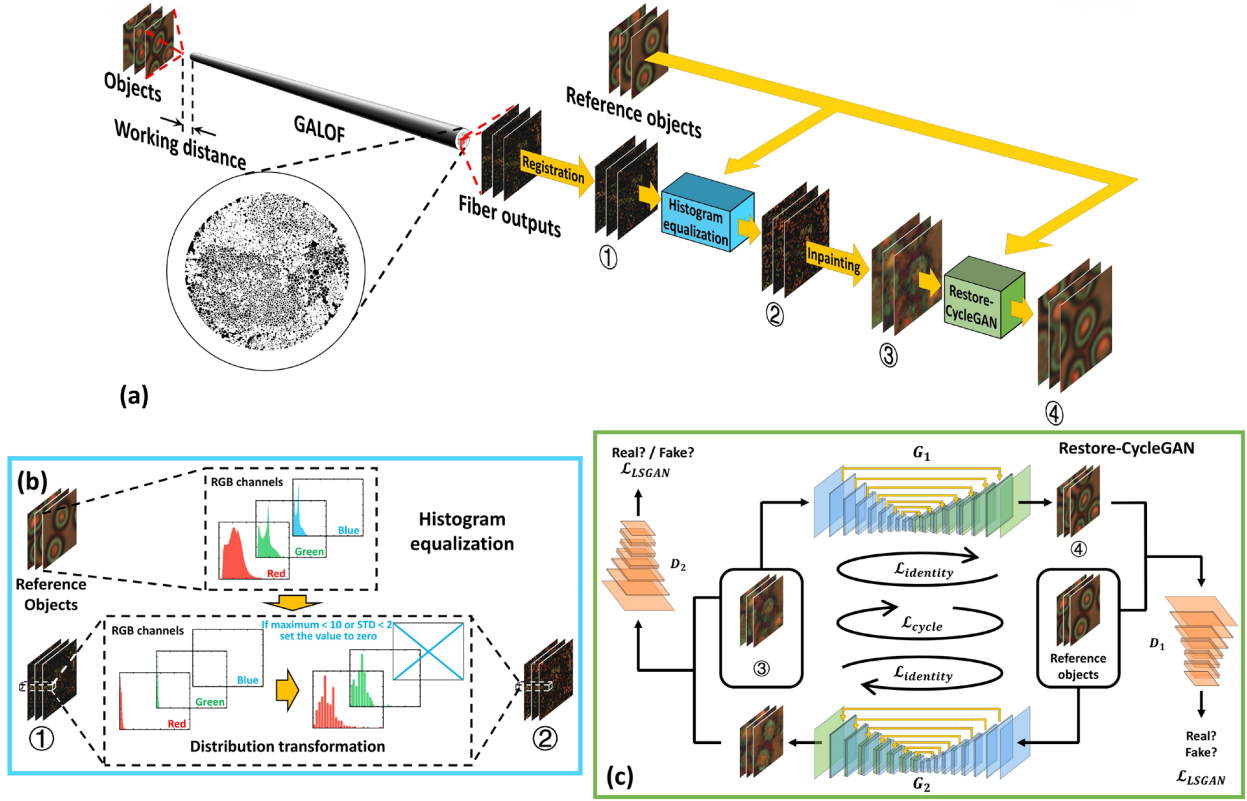

**Figure 1**. The calibration process of the unsupervised full-color image reconstruction. (a) Unknown objects are placed in front of the GALOF input facet at some working distance. To obtain the objects' statistical information, we separately collect unpaired reference objects. The calibration process consists of several steps. First, the GALOF outputs are registered (images ①) according to an arbitrarily chosen fiber output image. (b) The histogram equalization step. The PMF of each pixel in ① from (a) is transformed to resemble the PMF of the pixels in the reference objects. For each RGB channel, a LUT is created by matching the pixel values with the same CDF. "Bad" pixels with maxima less than 10 or STDs less than 2 are set to zero (e.g., blue channel in (b)), resulting in the images ② in (a). Then, the inpainting is performed on each image to fill in those bad pixels (the images ③ in (a)). (c) A Restore-CycleGAN recovers the image details. Two U-Net generators $G_1$ and $G_2$ translate between the images ③ and the reference object images, whereas two PatchGAN discriminators $D_1$ and $D_2$ distinguish the "real" images in the target domain from the "fake" generated images. Both the generators and the discriminators are optimized through the least square adversarial losses $L_{LSGAN}$. The generators are also updated through the identity mapping loss $L_{identity}$ and the cycle-consistent loss $L_{cycle}$. $L_{identity}$ requires an identical output if the input is in the target domain, while $L_{cycle}$ requires an unchanged image if the image goes through a full cycle. After learning, high-quality images ④ can be reconstructed by $G_1$.

When propagating through a GALOF (**Fig. 1 (a)**), the imaging information of the object is encoded by the TAL-supported modes. The light confinement provided by the TAL results in a nearly point-to-point transmission from the GALOF's input facet to the output facet. Due to the different mode losses, the GALOF output pattern is an unevenly weighted superposition of the TAL-supported modes. Reconstructing the object from the output pattern involves standardizing all the TAL-supported modes and solving an inverse imaging problem. This is a challenging task since the TAL-supported modes have a very high mode density. Instead, we tackle this problem by standardizing the pixels of the GALOF's output images. In the calibration step (**Fig. 1 (a)**), we collect 1,000 fiber output images and another unpaired 1,000 objects' reference images (**Methods**). Especially, there does not exist a one-to-one correspondence between these unpaired data sets. Before standardization, we register the 1,000 fiber outputs according to some arbitrarily chosen fiber outputs (**Methods**) to compensate for the image drift caused by the mechanical instability during experiments. As a result, each pixel in the fiber output has 1,000 different values. Statistically speaking, all these pixels should have the same Probability Mass Function (PMF) as those in the reference objects, despite being from different unknown objects. Therefore, we perform histogram equalization to each pixel in the fiber output image for each RGB (red, green, and blue) channel (**Fig. 1 (b)**). We calculate the Cumulative Density Function (CDF) from the PMF of each pixel and look for the pixel value in the reference objects that has the same CDF. In this way, we generate a Look-Up Table (LUT) for each pixel to transform its value. Among the range of 0 to 255, zeros are assigned to the defective pixels whose maximum value is less than 10 or whose Standard Deviation (STD) value is less than 2. For example, the value of the blue channel is set to zeros, as shown in **Fig. 1 (b)**. Next, we perform image inpainting (**Methods**) on each processed image. For each RGB channel, we interpolate inward from the pixels whose values are less than 10. Fuzzy objects are recovered after inpainting.

Finally, the reference objects are used again to further enhance the imaging quality of the 1,000 fuzzy objects. We utilize our recently proposed image restoration cycle-consistent adversarial network (Restore-CycleGAN) [23] (**Fig. 1 (c)**). As shown in our previous work, Restore-CycleGAN exhibits enhanced performance than the original CycleGAN [22] in extracting global information. In the Restore-CycleGAN, a U-Net [46] works as a generator $G_1$ to transform the fuzzy images into high-quality images, while a PatchGAN [47] works as a discriminator $D_1$ to distinguish the "real" reference objects from the "fake" ones produced by the generator $G_1$. The generator $G_1$ and the discriminator $D_1$ compete in an adversarial game through the least square adversarial loss $L_{LSGAN}$. In this adversarial game, $G_1$ gets rewarded if it successfully "fools" $D_1$, whereas $D_1$ gets rewarded if it differentiates the "real" from the "fake". $G_1$ is also optimized through the identity mapping loss $L_{identity}$, which requires a reference object to remain identical if it passes through $G_1$. Similarly, there is another pair of generator $G_2$ and discriminator $D_2$ in the opposite direction. To enhance cycle consistency, a cycle-consistent loss $L_{cycle}$ is adopted to enforce an unaltered output if an image goes through the two generators successively. The details of the network architectures and the training processes can be found in the **Methods**. After training, only the generator $G_1$ is used. Therefore, unsupervised image reconstruction is achieved without paired training data. During the test, a fiber output image goes through the process of: (1) aligning with the arbitrarily chosen fiber output image; (2) pixel value transformation using the LUTs; (3) inpainting; and (4) quality enhancement through the generator $G_1$. The set of reference objects is only needed to recalibrate the system for special cases, such as changing the working distances (**Fig. 1 (a)**).

## Results

### High fidelity

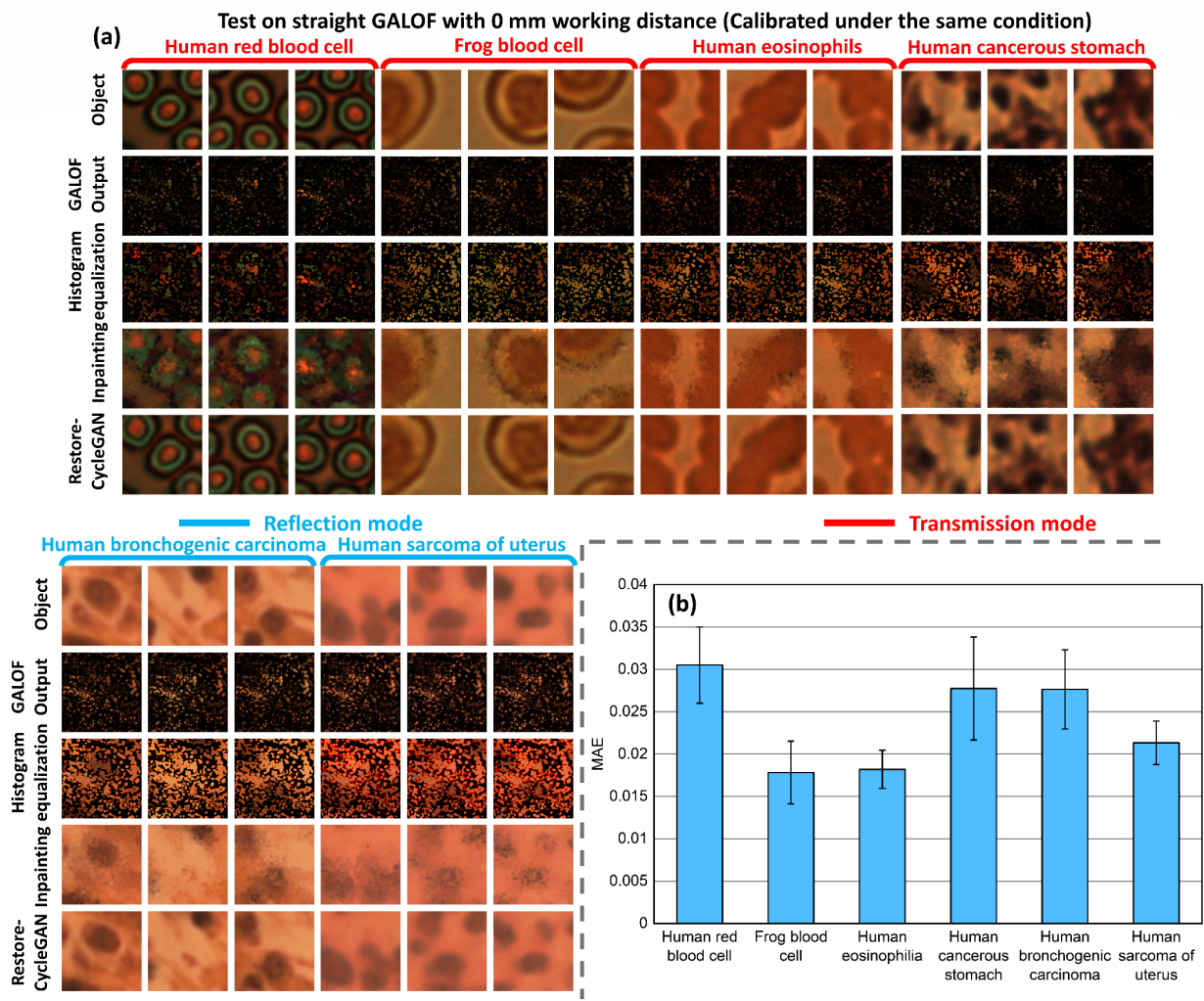

**Figure 2**. Test results of unsupervised full-color image reconstruction on different types of biological objects. (a) Sample images of the objects, the intermediate outputs of the reconstructions (excluding the registration step), and the final reconstructed images. The objects are human red blood cells, frog blood cells, human eosinophils, human cancerous stomach tissues, human bronchogenic carcinoma tissues, and human sarcoma of uterus tissues. The reconstruction process is calibrated and tested using the images collected through a straight GALOF with 0 mm working distance. The GALOF-based imaging system is in the transmission mode for the first four cases (the red brackets) and in the reflection mode for the last two cases (the blue brackets). (b) MAEs and STDs of the reconstructions are evaluated on 1,000 objects for each type of biological object.

We perform the calibration processes for six different biological objects: human red blood cells, frog blood cells, human eosinophils, human cancerous stomach tissues, human bronchogenic carcinoma tissues, and human sarcoma of uterus tissues. All calibrations are performed using the same straight GALOF with a working distance of 0 mm. As shown in **Fig. 2 (a)**, the data of the first four objects (human red blood cells, frog blood cells, human eosinophils, and human cancerous stomach tissues) are collected in the transmission mode, whereas that of the last two (human bronchogenic carcinoma tissues and human sarcoma of uterus tissues) are collected in the reflection mode. For each case, we separately collect 1,000 object images and their GALOF outputs to evaluate the performance. The reconstruction time per image is about 1.6 seconds. **Fig. 2 (a)** shows some examples of the objects' reference images, the GALOF output images, and the results after each reconstruction step (without the registration step). Although the raw GALOF outputs are unrecognizable, they preserve the local information of the objects well at all RGB channels. This is made clear after the histogram equalization is applied to the pixels in the registered images. After the inpainting step, fuzzy images of the objects start to show up. Finally, Restore-CycleGANs further recover the fine details. The high fidelity of the reconstructions is quantitatively demonstrated in **Fig. 2 (b)**, where we plot the mean absolute errors (MAEs) and STDs of the 1,000 reconstructions. In all six cases, the MAEs are below 0.035 (maximum ~1).

**Robustness**

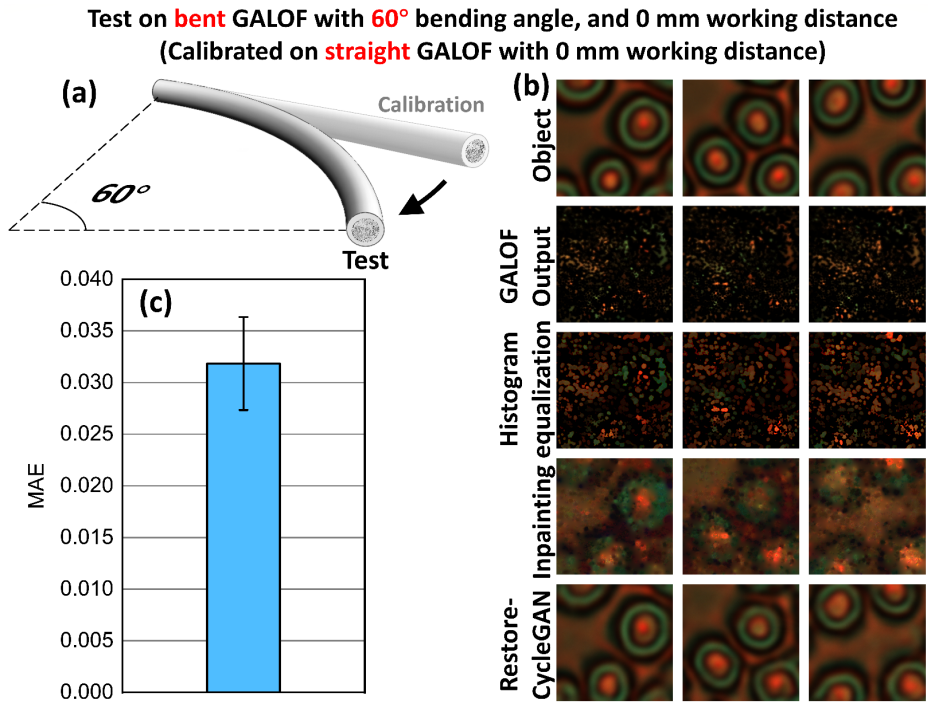

**Figure 3**. Robustness of the unsupervised image reconstruction under mechanical deformation. (a) After calibrating the LUTs and Restore-CycleGAN on human red blood cell samples through a straight GALOF with 0 mm working distance, we bend the GALOF with a centric angle of 60° and test the image reconstruction. (b) Sample images of the objects, the corresponding GALOF outputs, and the reconstruction. (c) MAE averaged over 1,000 reconstructions. The error bar stands for STD.

To test the robustness of the unsupervised fiber-optic imaging, we bend the GALOF with a central angle of 60° (**Fig. 3 (a)**). The output images are reconstructed using the LUTs and the Restore-CycleGAN calibrated on a straight GALOF. Both the calibration and test stages use the same type of human red blood cell samples at a working distance of 0 mm. As illustrated in **Fig. 3 (b)**, high-fidelity reconstructions are achieved despite the large-angle fiber bending. We repeat the image reconstruction for 1,000 GALOF outputs and calculate their MAE and STD with respect to the input objects (Fig. 3 (c)). Similar values can be observed from the test results on a straight fiber. This indicates the consistency of the GALOF outputs, as attributed to the excellent light confinement of the TAL-supported modes.

**Flexible working distance**

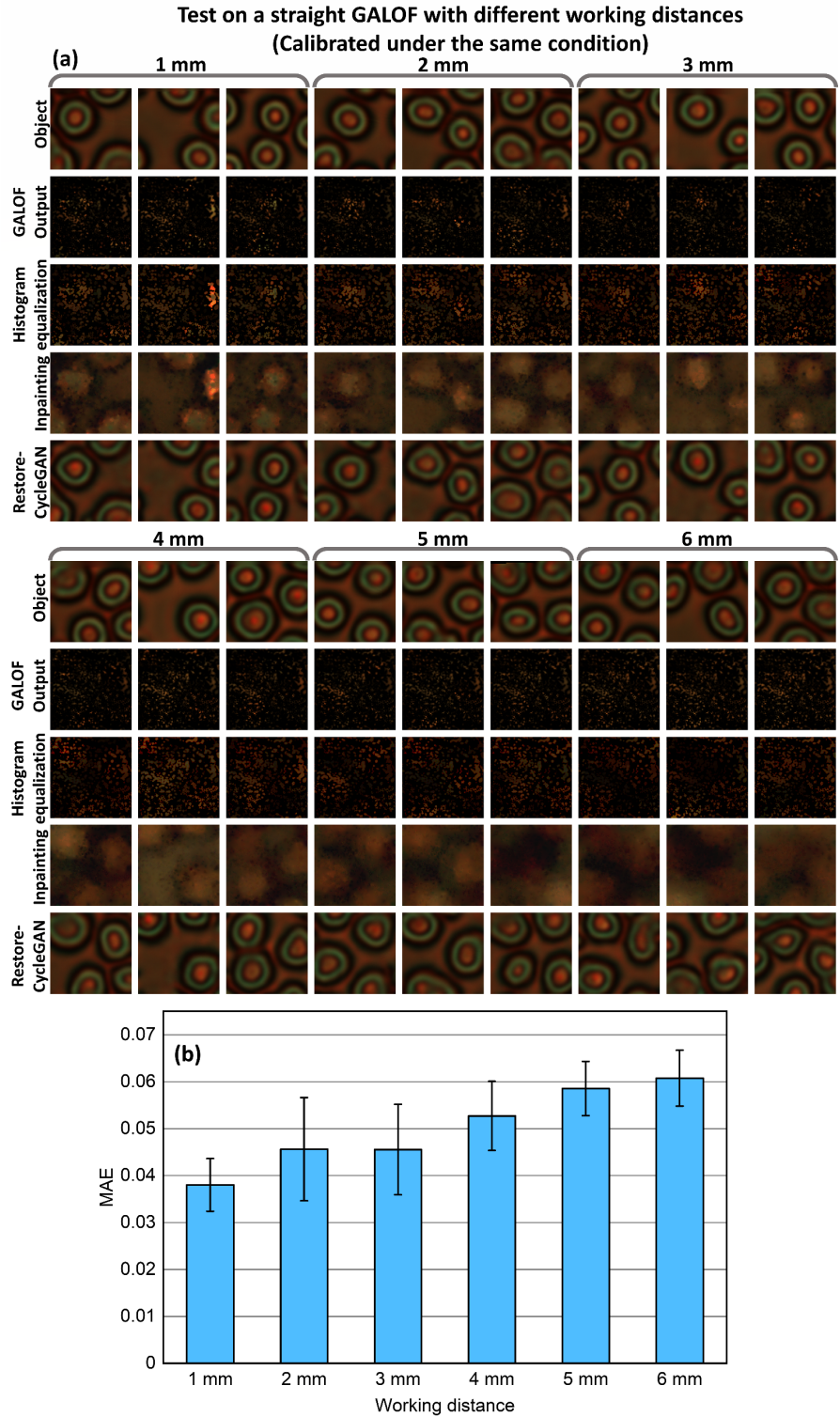

**Figure 4.** Test results of unsupervised full-color image reconstruction with different working distances. The reconstruction process is calibrated and tested in a straight GALOF at each working distance. (a) Sample images of the objects (human red blood cells), the outputs of each reconstruction step (excluding the registration step), and the final reconstructed images. (b) MAEs and STDs of the reconstructions evaluated on 1,000 objects for each type of biological objects.

Benefiting from the unsupervised image reconstruction, the reference objects only need to be collected once. When the working distance varies, we re-calibrate the LUTs and the Restore-CycleGAN using the same reference objects collected under a working distance of 0 mm. **Fig. 4** shows the test results on human red blood cells under the working distance of 1 mm to 6 mm with a step of 1 mm. Due to the loss of high-frequency information over the distance, the processed images after inpainting only demonstrate blurry profiles of the objects (**Fig. 4 (a)**). Nevertheless, Restore-CycleGANs can still recover the images of objects with fine details. High-fidelity reconstructions are preserved up to a working distance of at least 4 mm. With increased working distances, the processed images after inpainting lost more information with significantly degraded imaging qualities, resulting in false blood cell reconstructions by the Restore-CycleGANs, such as the reconstructions obtained at 6 mm working distance in **Fig 4 (a)**. To quantify the imaging performance, we calculate the MAEs and STDs of 1,000 pairs of reconstructions and ground truths at each working distance (**Fig. 4 (b)**). It shows that the increasing working distance does not rapidly degrade the image quality. Therefore, our unsupervised image reconstruction approach enables flexible working distances through a simple re-calibration procedure.

**Cross-domain generalizability**

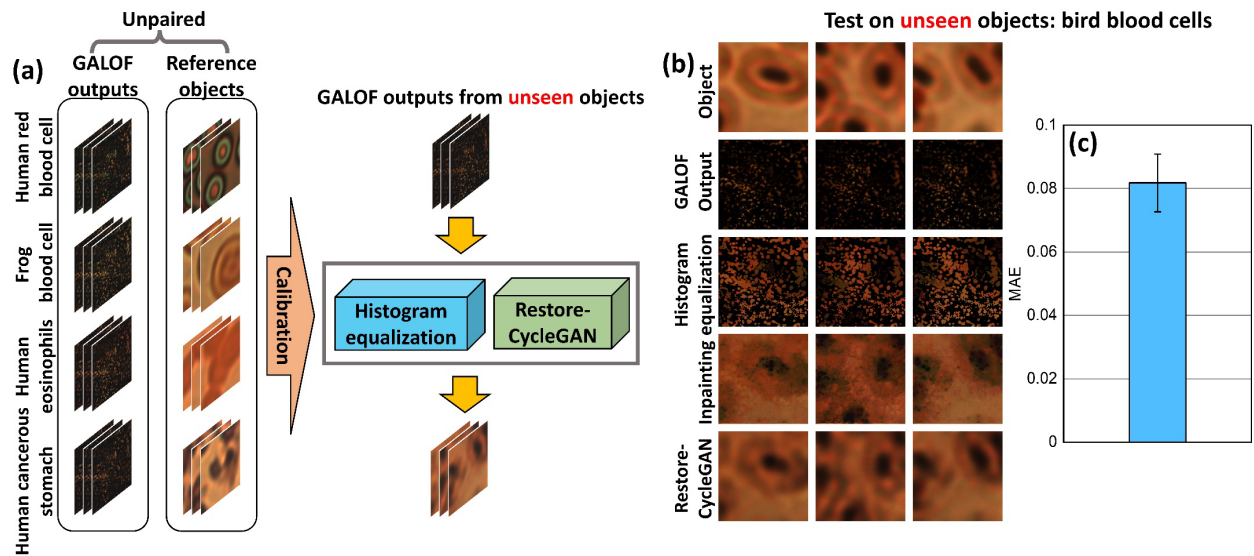

**Figure 5.** Cross-domain generalizability on unseen objects. (a) In the calibration, both the GALOF outputs and the reference objects contain mixed images from four different types: human red blood cells, frog blood cells, human eosinophils, and human cancerous stomach tissues. We collect 300 unpaired and uncorrelated images for each type of sample from the GALOF outputs and the reference objects. Following the same procedure (**Fig. 1**), we obtain the LUTs for histogram equalization and the Restore-CycleGAN. Then we test the image reconstruction on GALOF outputs from unseen objects, i.e., bird blood cells. (b) Sample images of the bird blood cell images, the corresponding GALOF outputs, and the outputs of reconstruction. (c) MAE and STD averaged over 300 reconstructions.

In the abovementioned results, the calibration and test are conducted on the same type of biological objects. Yet, FOISs are expected to perform high-fidelity imaging on unseen objects in real-world applications. To enhance cross-domain generalizability, it is necessary to enrich the statistical information of the objects for image reconstruction. For this purpose, we include image data generated from various types of biological samples (**Fig. 5 (a)**), such as human red blood cells, frog blood cells, human eosinophils, and cancerous stomach tissues. For each object type, we collect 300 reference object images and 300 GALOF output images from a straight GALOF with 0 mm working distance in the transmission mode. These two sets of images are unpaired and uncorrelated. We follow the same calibration procedure demonstrated in **Fig. 1**. After obtaining the LUTs and the Restore-CycleGAN, we test the unsupervised image reconstruction on GALOF outputs from unseen objects, i.e., bird blood cells. **Fig.5 (b)** shows sample object images, the corresponding GALOF outputs, and the processed images after each reconstruction step (excluding the registration step). The profiles and orientations of the bird blood cells can be clearly identified, despite slightly degraded image quality. This corresponds to a higher MAE over 300 reconstructed images (**Fig. 5 (c)**). The increase in MAE originates from the limited data size and object variations, which could be addressed by improving the training data further.

## Discussion

Robust full-color high-fidelity image transport using unsupervised learning is achieved through the unique properties of GALOFs. First, the high-density localized modes result in a point-to-point transmission of the object with a high sampling ratio, which lays the foundation of unsupervised image reconstruction. Second, the TAL-supported modes have flat responses to different wavelengths [38], enabling full-color imaging. In contrast, an optical fiber bundle designed to transmit images at one wavelength may not be suitable for another wavelength[13]. Third, the robustness of the TAL-supported modes makes the imaging performance stable under large fiber deformations, whereas a translation of a few millimeters in one end of a meter-long optical fiber bundle or MMF tends to significantly degrade the output image [24, 41, 42, 43].

Unsupervised image reconstruction simplifies the calibration process and makes the GALOF-based FOISs flexible towards different circumstances. For example, a wide range of working distances is desirable in endoscopy applications to reduce penetration damage. System re-calibration by supervised learning is impractical since it requires collecting paired images. As shown in this work, unsupervised learning enables simple re-calibrations of the system to acquire high-quality images up to a working distance of at least 4 mm. Moreover, the amount of data needed in calibrations is dramatically reduced. In our experiments, we only acquire 1,000 GALOF outputs and 1,000 reference object images for one calibration. In contrast, supervised learning typically requires tens of thousands of paired images to train a CNN.

It is worth noting that reconstructions by the Restore-CycleGAN cannot be valid for arbitrary scenarios. The two image domains should not deviate too much from an identical mapping so that the Restore-CycleGAN can find a 'natural' translation [23]. More generally, it has been shown that unsupervised image-to-image translation often fails when a more extreme transformation takes place between the two image domains [22]. Therefore, registration, standardization, and inpainting steps are essential to high-fidelity reconstructions using the Restore-CycleGAN in bringing the GALOF outputs close to the object image domain.

Further improvement can be made in the GALOF fabrications and the unsupervised image reconstruction process. Currently, there are many defective pixels in the GALOF outputs, which lead to the loss of information. Future work can be devoted to investigating methods of eliminating these defective pixels. Moreover, the geometrical parameters of the GALOF can also be improved. The GALOF used in this work has an air-filling fraction of ~28%. In contrast, an air-filling fraction of ~50% has been shown to be favorable for reducing the localization lengths and improving spatial resolution [48, 49]. On the other hand, there is still much room for improvements in cross-domain generalizability. We expect a larger and more diversified image dataset would enhance the image reconstruction of unseen objects in future work. Furthermore, while it only takes the

Restore-CycleGAN about 70 ms to reconstruct an image, the steps preceding the Restore-CycleGAN add a significant amount of time to the whole reconstruction process. The reconstruction time per image is about 1.6 seconds. Future studies can focus on speeding up the processing speed for steps preceding the Restore-CycleGAN to realize real-time imaging.

In conclusion, we achieve unsupervised image reconstruction in a meter-long GALOF based on its unique property of point-to-point transmission with high sampling densities. Full-color high-fidelity image transport is demonstrated on different types of biological samples in both transmission and reflection modes. The image quality is preserved when the GALOF is substantially bent with an angle of 60°. Enabled by unsupervised image reconstruction, the GALOF-based FOIS is flexible to different circumstances. High image quality is maintained within a working distance of at least 4 mm using a much-simplified re-calibration. Increased cross-domain generalizability on unseen objects is also shown by including diversified objects. Based on these results, we see the GALOF-based FOISs as promising candidates for the next-generation FOISs.

## Methods

### Experimental setup

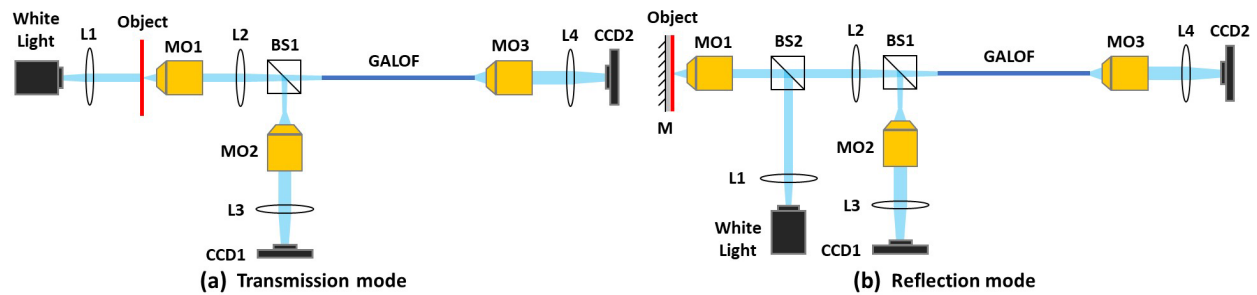

**Extended Data Figure 1.** (a) Experimental setup under the transmission mode. (b) Experimental setup under the reflection mode. L1: collimating lens. L2, L3, L4: tube lenses. MO1, MO2, MO3: microscope objectives. BS1, BS2: beam splitters. M: reflective mirror. In the transmission mode, the white light illuminates the biological object from the back. In the reflection mode, the sample illumination and the image acquisition use the same objective, MO1. In both modes, the object image is relayed by MO1 and L2, split by BS1 into two copies. Each copy transmits through the imaging arm GALOF-MO3-L4-CCD2 and the reference arm MO2-L3-CCD1. The two arms operate separately to collect unpair images for calibration, while both arms operate synchronously to collect unpaired images for the test.

In both the transmission mode and the reflection mode (**Extended Data Fig. 1 (a), (b)**), we use a quartz halogen lamp as the light source (wavelength: ~400nm to ~2000nm). A lens, L1, is placed in front of the lamp to collimate the light. In the transmission mode, the collimated light illuminates the object from behind. The object image is relayed by a 10x microscope objective (MO1) (infinity-corrected, NA = 0.3) and a tube lens L2 (f = 200 mm). The magnified image is then sent to two arms by a beam splitter BS1. In the reference arm, the image is further magnified by a 20x microscope objective (MO2) (NA = 0.75) and a tube lens L3 (f = 200 mm), and then collected by the CCD1 camera (Manta G-145C). In the imaging arm, the object image is delivered through the GALOF. The GALOF is fabricated by the stack-and-draw method. It has a disordered structure with a diameter of ~278 μm and an air-hole-filling fraction of ~28.5%. A segment of ~80 cm is used in the experiment. The GALOF output is magnified by a combination of a 20x microscope objective (MO3) (NA = 0.75) and a tube lens L4 (f = 200 mm) before being collected by the CCD2 camera (Manta G-145C). In the reflection mode, the illumination light is coupled into the back aperture of the MO1 by a beam splitter (BS2), and focused onto the object. We place a mirror M as a highly reflective substrate behind the object without contact. Similar to the transmission mode, the reflected object image is magnified and sent to the two imaging arms. For both the transmission and reflection modes, the reference arm and the imaging arm collect images separately during calibration. They only operate synchronously during the test to evaluate the system's imaging performance.

## GALOF outputs registration and inpainting

We first convert all the GALOF outputs to grayscale images. To find the transformation for registration, we use MATLAB "*imregtform*" function with monomodal registration and translation geometric transformation. Inpainting is based on MATLAB's "*regionfill*" function that can interpolates inward from the values of the pixels on the outer boundary of the regions.

## Restore-CycleGAN

The architectures of the generator and discriminator networks in the Restore-CycleGAN are shown in **Extended Data Fig. 2 (a), (b)**. The generator is a U-Net with skip connections. The input image has a size of 420×420. It passes through different convolutional layers to a bottleneck and then passes through different transposed convolutional layers to the final output. The discriminator is a PatchGAN, which looks into patches of an input image and decides whether they are from the real images in the target domain or from the fake images generated by the generator. The detailed operations in the convolutional layers and transpose convolutional layers are shown in **Extended Data Fig. 2 (c)**. The numbers of filters in the layers of the generator are 64-128-256-512-512-512-512-512-512-512-512-512-256-128-64-3. The numbers of filters in the layers of the discriminator are 64-128-256-512-512-1.

The weights in the generators and the discriminators are initialized by random Gaussian distributions with a zero mean and a standard deviation of 0.02. For translating between the domain $x$ and the domain $y$, there are two generator-discriminator pairs: $G_{x \to y}$ and $D_y$ in the direction from $x$ to $y$, and $G_{y \to x}$ and $D_x$ in the other direction. The loss function of a generator $G_{x \to y}$ can be written as:

$$\begin{aligned}
\mathcal{L}_{G_{x \to y}} = &\, \mathrm{E}_x \left[ \left( D_y \left( G_{x \to y}(x) \right) - 1 \right)^2 \right] \\
&+ \alpha_1 \mathrm{E}_y \left[ \left\| G_{x \to y} \left( G_{y \to x}(y) \right) - y \right\|_1 \right] \\
&+ \alpha_1 \mathrm{E}_x \left[ \left\| G_{y \to x} \left( G_{x \to y}(x) \right) - x \right\|_1 \right] \\
&+ \alpha_2 \mathrm{E}_y \left[ \left\| G_{x \to y}(y) - y \right\|_1 \right]
\end{aligned} \quad (1)$$

The four terms in Eq. (1) are the least square adversarial loss $\mathcal{L}_{LSGAN}$, the cycle-consistent losses $\mathcal{L}_{cycle}$ in both directions, and the identity mapping loss $\mathcal{L}_{identiy}$, respectively. $\alpha_1$ and $\alpha_2$ are the weights controlling the balance among the losses. We use $\alpha_1 = 10$, and $\alpha_2 = 5$. The weights in $D_y$ and $G_{y \to x}$ are fixed when we train $G_{x \to y}$. The loss function of $D_y$ is the least square adversarial loss $\mathcal{L}_{LSGAN}$:

$$\mathcal{L}_{D_y} = \mathrm{E}_y \left[ \left( D_y(y) - 1 \right)^2 \right] + \mathrm{E}_x \left[ D_y \left( G_{x \to y}(x) \right)^2 \right] \quad (2)$$

The weights in $G_{x \to y}$ are fixed when we train $D_y$. The real images $y$ to train $D_y$ are randomly selected from all the images in the target domain, whereas the fake images $G_{x \to y}(x)$ are randomly selected from a pool of 50 fake images. The pool is randomly updated through newly generated fake images. The loss of the discriminators is divided by half during training. The loss functions of $G_{y \to x}$ and $D_x$ can be written in a similar way. We train the discriminators and generators for 100 epochs with a batch size of 1. We use an Adam optimizer with a learning rate of 0.0002 and the exponential decay rate for the first momentum β₁=0.5. The training takes ~40 hours on a dual-GPU (GeForce GTX 1080 Ti) desktop.

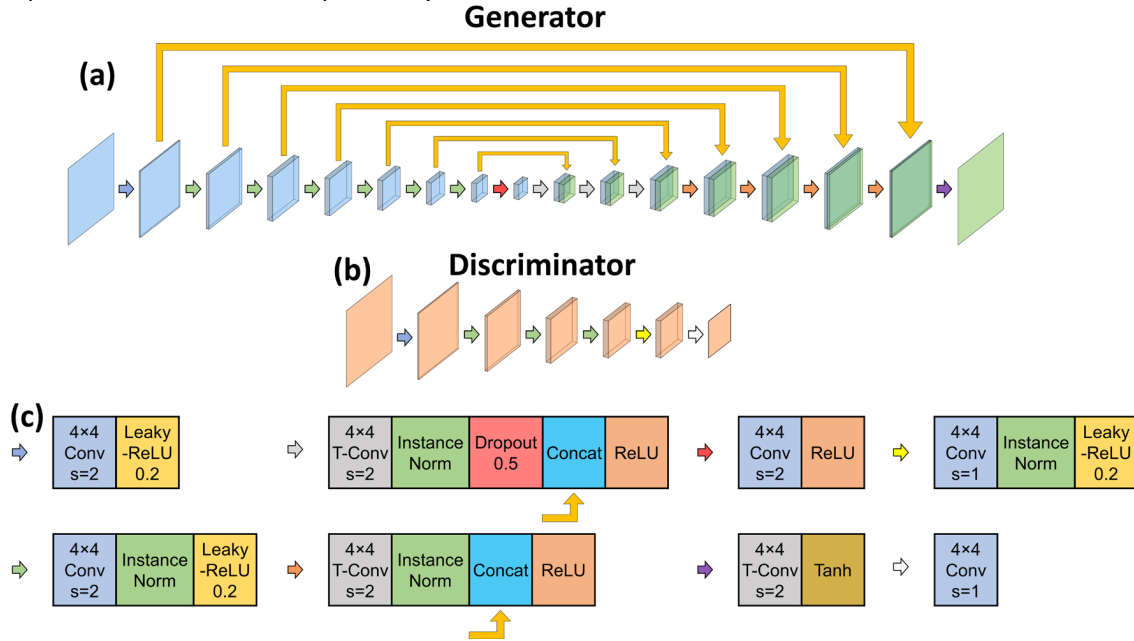

**Extended Data Figure 2.** (a) The generator architecture in Restore-CycleGAN. (b) The discriminator architecture in Restore-CycleGAN. (c) The detailed explanation of the arrows in (a) and (b). Conv: convolution. T-Conv: transpose convolution. InstanceNorm: instance normalization. s: stride.

## Data availability

The datasets generated during the current study are available from the authors under reasonable request.

## Code availability

The codes developed for this work are available from the authors under reasonable request.

## Author Contributions

X.H. developed the GALOF-based unsupervised full-color imaging framework, including the algorithms and the experimental systems, performed the data acquisitions, processed the experimental data, and wrote the first draft. X.H. and J.Z. proposed this project. J.Z. developed the prototype of the transmission imaging system, supervised the project, and revised the manuscript. J.Z. and J.E.A. developed and fabricated the GALOF used in the experiment. R.A.C. and A.S. supervised the GALOF development and fabrication. A.S. led the team, supervised the project, and revised the manuscript. All authors contributed to the final version of the manuscript.

## Conflict of interest

The authors declare no conflict of interest.